\documentclass[aps,prb,twocolumn,showpacs,amsmath,amssymb]{revtex4}
\usepackage{graphicx}
\usepackage{dcolumn}
\usepackage{bm}

\begin{document}
\title{First-principles investigation of symmetric and antisymmetric exchange interactions of SrCu$_{2}$(BO$_{3}$)$_{2}$}
\author{V. V. Mazurenko$^{1,3}$, S.L. Skornyakov$^{1}$, V. I. Anisimov$^{1,2}$, F. Mila$^{3}$  }
\affiliation{$^{1}$Theoretical Physics and Applied Mathematics Department, Urals State Technical University, Mira Street 19,  620002
Ekaterinburg, Russia \\
$^{2}$Institute of Metal Physics, Russian Academy of Sciences, 620219 Ekaterinburg GSP-170, Russia \\
$^{3}$Institute of Theoretical Physics, Swiss Federal Institute of Technology (EPFL), CH-1015 Lausanne, Switzerland}
\date{\today}

\begin{abstract}
We report on a first-principles investigation of the electronic structure and of the magnetic properties of the quasi-two-dimensional Mott insulator SrCu$_{2}$(BO$_{3}$)$_{2}$.
Based on the hopping integrals and Coulomb interactions calculated with LDA and LSDA+U, we provide a microscopic explanation of the symmetric Heisenberg and antisymmetric Dzyaloshinskii-Moriya exchange integrals of SrCu$_{2}$(BO$_{3}$)$_{2}$. 
The intra-atomic exchange interaction of oxygen is shown to strongly contribute to the intra-dimer isotropic exchange.
The results are in good agreement with those derived from experimental data, both regarding the orientation of the Dzyaloshinskii-Moriya vectors and the magnitude of all exchange integrals. The 
microscopic analysis is confirmed by the results of Green function's and total energies 
difference methods.
\end{abstract}

\maketitle
\section{Introduction}
The quasi-two-dimensional compound SrCu$_{2}$(BO$_{3}$)$_{2}$ has attracted a lot of attention over the past ten years due to its spin-gapped behavior, its finite magnetization below the critical magnetic field deduced from the spin gap, and more importantly its magnetization plateaus at 1/8, 1/4 and 1/3 of the saturated magnetization. 
\cite{Miyahara,Kageyama,Onizuka}
Based on the quasi-two-dimensional structure of the compound and on a number of experimental facts, the Hamiltonian expected to provide an accurate description of the magnetic properties of that compound must include both 
Heisenberg symmetric exchange processes and antisymmetric Dzyaloshinskii-Moriya (DM) interactions. It is defined by:
\begin{eqnarray}
H=J\sum_{n.n} \vec S_{i} \vec S_{j} + J' \sum_{n.n.n} \vec S_{i} \vec S_{j} \nonumber \\
+ \vec D \sum_{n.n} [\vec S_{i} \times \vec S_{j}] + \vec D' \sum_{n.n.n} [\vec S_{i} \times \vec S_{j}], 
\end{eqnarray} 
where J ($\vec D$) and $J'$ ($\vec D'$) define intra- and inter-dimer symmetric (Dzyaloshinskii-Moriya) exchange interactions. The Heisenberg model obtained when $\vec D$ and $\vec D'$ are set to zero is known as the Shastry-Sutherland model\cite{shastry}. The susceptibility and the main features of the magnetization curve have been interpreted in the context of this Shastry-Sutherland model, leading to estimates of $J$ and $J'$, while
the dispersion of triplet excitations, the low-field uniform and staggered magnetizations and ESR measurements
have provided estimates of the DM interactions. A very useful complementary source of information is usually
given by first-principles calculations,
which have provided valuable results even in low-dimensional quantum spin systems with complex structures. \cite{mazurenkoLiCu2O2,mazurenkoNa2V3O7}
Surprisingly enough, no ab-initio results have been reported so far for SrCu$_2$(BO$_3$)$_2$.

In this paper, we report on the first ab-initio investigation of the electronic structure and magnetic properties of SrCu$_{2}$(BO$_{3}$)$_{2}$. 
Our results show that an accurate treatment of the oxygen magnetization is crucial
for the description of isotropic exchange interactions in SrCu$_{2}$(BO$_{3}$)$_{2}$.
In order to calculate the Dzyaloshinskii-Moriya interactions between magnetic moments,
we have used the microscopic expression derived by Moriya.\cite{Moriya} For that purpose, the hopping integrals between different Wannier orbitals centered at Cu atoms have been obtained using a projection procedure.

The paper is organized as follows: in the next section, we shortly describe
the crystal structure of SrCu$_{2}$(BO$_{3}$)$_{2}$ and present the results of LDA calculations. 
Sec.III and IV contain the analysis of the isotropic and anisotropic exchange interactions, The results of LSDA+U calculations are presented in Sec.V. We discuss and briefly summarize our results in Sec.VI and VII .

\section{Results of LDA calculations}
The simplified crystal structure of SrCu$_{2}$(BO$_{3}$)$_{2}$ is presented in Fig.1.
Each copper atom has one nearest-neighbor Cu atom and four next-nearest-neighbor Cu atoms in CuBO$_{3}$ layers.
In Sr layers, the distance between Cu$^{2+}$ ions is shorter than in CuBO$_{3}$ layers. However, our results demonstrate, as we will show below, that the
magnetic interactions inside the CuBO$_{3}$ layers are much stronger than in the Sr layers. 
\begin{figure}[!h]
\centering
\includegraphics[width=0.4\textwidth]{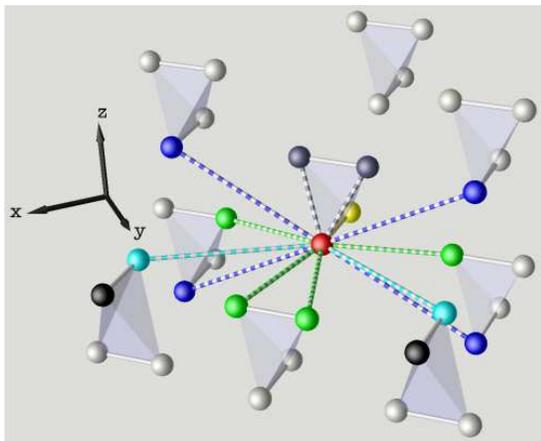}
\caption {Simplified three-dimensional structure of SrCu$_{2}$(BO$_{3}$)$_{2}$. 
The cycles are copper atoms. The dashed lines correspond
to inter-dimer interaction paths.}
\end{figure}

The electronic structure calculation of SrCu$_{2}$(BO$_{3}$)$_{2}$
has been performed using the Tight Binding Linear-Muffin-Tin-Orbital 
Atomic Sphere Approximation (TB-LMTO-ASA) method in terms of the 
conventional local-density approximation.\cite{Andersen}
We used the known crystal structure data.\cite{smith}  The radii of atomic spheres have been set to r(Cu) = 2.2 a.u.,
r(O) = 1.65 a.u., r(B) = 1.3 a.u., r(Sr) = 3.5 a.u. and nine types of empty spheres were added.

Fig.2 gives the total and partial density of states projected on the constitutional atoms of the SrCu$_{2}$(BO$_{3}$)$_{2}$.
The following characteristics can be seen from Fig.2. (1) The bands lower than - 15 eV mostly consist of 2s orbitals of the oxygen atoms of the (BO$_{3}$)$^{3-}$
complex. (2) The density of states near Fermi level comes mainly from Cu and O orbitals.
\begin{figure}[!b]
\centering
\includegraphics[angle=270,width=0.4\textwidth]{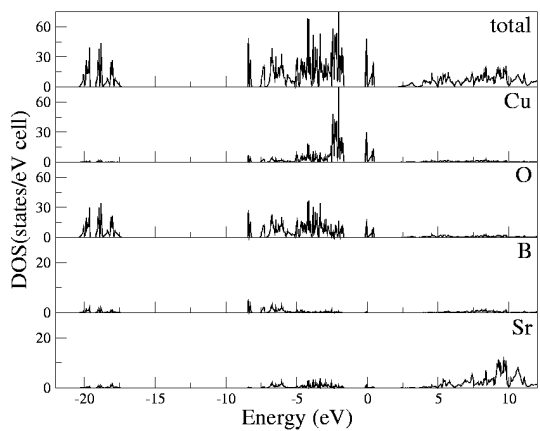}
\caption {Total and partial density of states of SrCu$_{2}$(BO$_{3}$)$_{2}$  obtained in LDA calculations.}
\end{figure} 
The band structure of SrCu$_{2}$(BO$_{3}$)$_{2}$ near the Fermi level obtained by LDA calculation is presented in Fig.3.
There are four well separated bands. The analysis of the partial density of states (Fig.4)
shows that the main contribution to these bands comes from copper orbitals with x$^{2}$-y$^{2}$ symmetry.  
\begin{figure}[!b]
\centering
\includegraphics[angle=270,width=0.45\textwidth]{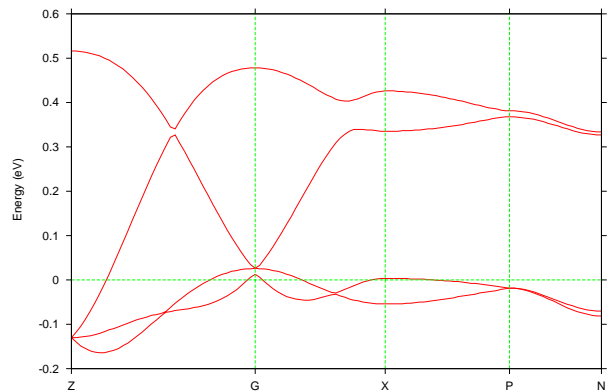}
\caption {Band structure of SrCu$_{2}$(BO$_{3}$)$_{2}$ near Fermi level (0 eV).}
\end{figure}
There is however a contribution of oxygen 2p states due to the strong hybridization of Cu 3d and O 2p states (see Fig.2 and Fig.4). 
In this situation, the most natural and simplest way to describe the magnetism and to take into account the hybridization between copper and oxygen is work in a
Wannier function basis. \cite{wannier,Pickett} 
We have defined the Wannier functions centered on the copper orbitals with $x^2-y^{2}$ symmetry using 
a projection procedure. \cite{proj} The resulting orbitals (presented in Fig.5) have a strong 
contribution from the oxygen atomic wave orbitals and can be expressed through the following linear combination:
\begin{eqnarray}
W_{i} (x) = \alpha \phi_{d_{x^2-y^2}} + 2 \beta (\phi_{p_x} + \phi_{p_y}),
\end{eqnarray}
where $\alpha$ and $\beta$ are the amplitudes of the copper and oxygen atomic wave functions in the Wannier orbital $W_{i} (x)$.
\begin{figure}[!t]
\centering
\includegraphics[angle=0,width=0.45\textwidth]{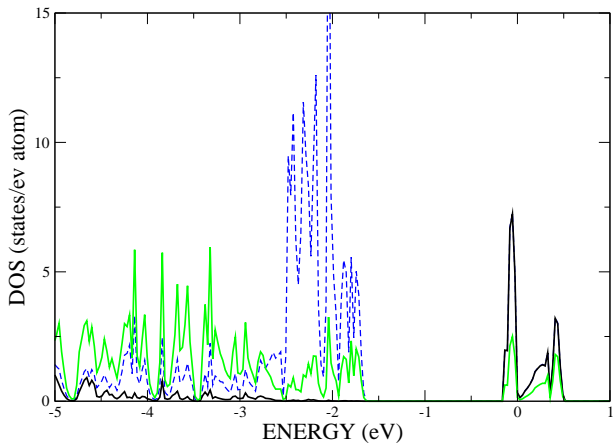}
\caption {Partial density of 3d copper states obtained by LDA calculations.
The blue dashed, black and green solid lines are 3d states of copper, 3d$_{x^{2}-y^{2}}$ orbital and 2p states of oxygen, respectively.
The zero energy corresponds to the Fermi level.}
\end{figure}

\begin{figure}[!b]
\centering
\includegraphics[angle=90,width=0.42\textwidth]{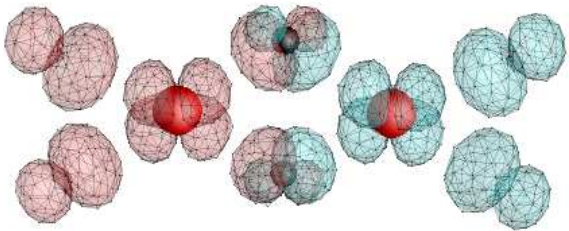}
\caption {Wannier orbitals centered at $x^2-y^2$ orbitals of copper 
atoms which belong to the dimer.}
\end{figure}

\section{Isotropic exchange interactions}
The results of the previous section demonstrate that within 
LDA the SrCu$_{2}$(BO$_{3}$)$_{2}$ has 
a metallic ground state. This is a standard LDA problem for transition
metal oxides and in order to overcome it one should take into account on-site 
correlations in the electronic density functional. 
Despite this problem, a microscopic description of magnetic couplings is still 
possible on the basis of LDA. In order to demonstrate this, we construct a model 
Hamiltonian describing SrCu$_{2}$(BO$_{3}$)$_{2}$ in the Wannier function basis. This Hamiltonian 
is set to reproduce only four bands of the
full Hamiltonian near the Fermi level (Fig.3).
We express the LDA structure 
results in terms of a low-energy, few orbital model Hamiltonian using a
projection procedure.\cite{proj} 
Let us first consider the one-orbital tight-binding Hamiltonian in the Wannier functions basis
$H_{TB}=-\sum_{i,j, \sigma} t^{x^2-y^2}_{ij} a_{i \sigma}^{+} a_{j \sigma}$, where 
$t^{x^2-y^2}_{ij}$ is the hopping integral between Wannier functions centered on the 3d$_{x^2-y^2}$ orbitals of {\it i}th and {\it j}th sites. 
To simplify the analysis we divide the calculated hopping integrals into two subgroups: intraplane (Fig.6) and interplane (Fig.7) couplings. The intraplane transfers are t$_{12}^{x^2-y^2}$= 169 meV, t$_{13}^{x^2-y^2}$ = t$_{14}^{x^2-y^2}$= t$_{15}^{x^2-y^2}$= t$_{16}^{x^2-y^2}$ = 63 meV, t$_{17}^{x^2-y^2}$=t$_{18}^{x^2-y^2}$=15 meV and t$_{19}^{x^2-y^2}$ = t$_{1 \, 10}^{x^2-y^2}$= t$_{1 \, 11}^{x^2-y^2}$= t$_{1 \, 12}^{x^2-y^2}$ = 12 meV. 
In turn the interplane hopping integrals have the values: t$_{1A}^{x^2-y^2}$=t$_{1B}^{x^2-y^2}$=8 meV and
t$_{1C}^{x^2-y^2}$ = t$_{1 D}^{x^2-y^2}$= t$_{1F}^{x^2-y^2}$= t$_{1E}^{x^2-y^2}$ = 12 meV.

\begin{figure}[!t]
\centering
\includegraphics[width=0.4\textwidth]{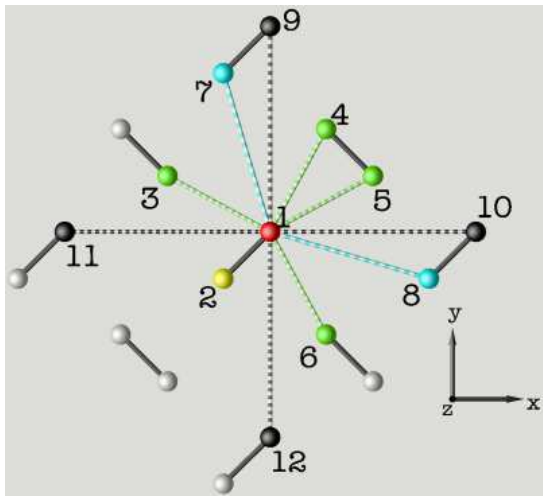}
\caption {Representation of couplings between Atom 1 and its neighbours which belong to the plane.
The interactions between Atom 1 and the light grey sites without numbers are negligibly small.}
\end{figure}

\begin{figure}[!b]
\centering
\includegraphics[width=0.4\textwidth]{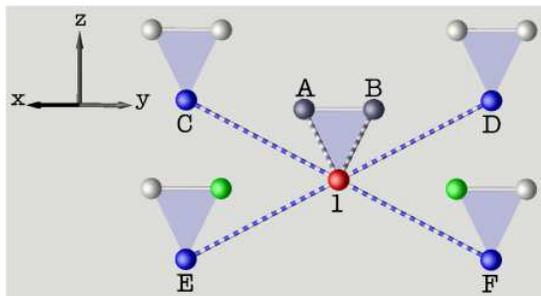}
\caption {Interplane interactions between copper atoms. The interactions between Atom 1 and the light grey sites without letters are negligibly small.}
\end{figure}
Using these hopping parameters, one can estimate the magnetic couplings of SrCu$_{2}$(BO$_{3}$)$_{2}$.
Since the Wannier orbitals have a strong contribution from the wave functions of the oxygen atoms, it is not enough to use the simple formula $\frac{4t^{2}_{ij}}{U}$. There is an additional ferromagnetic contribution originating from Hund's rule intra-atomic exchange interaction of the oxygen\cite{mazurenkoLiCu2O2}, which leads to the formula:
\begin{eqnarray} 
J = \frac {4 t^{2}_{ij}}{\alpha^{4} 
U_{d}} - 
2 \beta^{4} J^{H}_{p} N_{ox}, 
\end{eqnarray}
where $N_{ox}$ is the number of oxygen atoms between coppers and 
$\beta$ is the contribution of atomic wave functions of oxygen 
to the Wannier orbitals. U$_{d}$ and $J_{p}^{H}$ are on-site Coulomb interactions of copper atom and intra-atomic exchange interaction of oxygen, respectively. The former interaction can be estimated within constrained LDA, which
gives U$_{d}$=8.4 eV.
According to spin-polarized LSDA+U calculations \cite{mazurenkoLiCu2O2} $J^{H}_{p}$ =1.6 eV. 
The value of the copper magnetic moment obtained by LSDA+U calculations is 0.72 $\mu_{B}$.
One can estimate $\beta^{2}$ through the magnetic moment of the oxygen atom in the
ferromagnetic configuration simulated in LSDA+U calculations (Table III): $\beta^{2}$ = M(O)/2=0.05.
Using Eq.(3) and the parameters defined above leads to: $J_{12}$ = 10.2 meV. The value of $J_{12}$ is larger than that derived from experimental data. The problem might be that Eq.(3) is sensitive to the form of the Wannier functions. For instance, if $\beta^2$ was equal to 0.054 (instead of 0.05), then $J_{12}$ would be equal to 7.5 meV, in excellent agreement with the experimental value. A detailed comparison of our results for the magnetic couplings with those extracted from experiments is presented in section VI.

On the other hand, there is no overlap at oxygen atom between Wannier functions which belong to different dimers and therefore the inter-dimer interactions can be calculated using the standard expression: 
\begin{eqnarray}
J_{ij} = \frac{4 t^{2}_{ij}}{\alpha^{4} U_{d}}.
\end{eqnarray}
This leads to $J_{13}$=3.6 meV, $J_{17}$=0.2 meV, $J_{19}$=0.13 meV, $J_{1A}$ = 0.06 meV and $J_{1C}$ = 0.13 meV. 

\section{Anisotropic exchange interactions}
In a pioneer investigation, Moriya \cite{Moriya} pointed out two main contributions to the anisotropic exchange interaction. The first one is the kinetic Dzyaloshinskii-Moriya interaction, which can be expressed in the following form:
\begin{eqnarray}
\vec D^{kin}_{ij} = \frac{8i}{\alpha^{4} U_{d}} [t^{nn'}_{ij} \vec C_{ji}^{n'n} - \vec C_{ij}^{nn'}t_{ji}^{n'n}],
\end{eqnarray}
where $n$ and $n'$ denote the ground state Wannier orbitals centered at the {\it i}th and {\it j}th sites, while $t^{nn'}_{ij}$ and $\vec C_{ij}^{nn'}$ are transfer integrals without and with spin-orbit coupling. In the case of SrCu$_2$(BO$_3$)$_2$, the ground state Wannier function is centered on the  
3d$_{x^2-y^2}$ orbital of copper.
If the energy difference between the ground state and the excited states is larger than the spin-orbit coupling, then we can treat the spin-orbit interaction as a perturbation. In this case $\vec C_{ji}^{n'n}$ is given by
\begin{eqnarray}
\vec C_{ji}^{n'n} = -\frac{\lambda}{2}(\sum_{m'} \frac{\vec l^{m'n' *}_{j}}{\epsilon_{j}^{m'}-\epsilon_{j}^{n'}}
t_{ji}^{m'n} + \sum_{m} \frac{\vec l^{mn}_{i}}{\epsilon_{i}^{m}-\epsilon_{i}^{n}}
t_{ji}^{n'm}),
\end{eqnarray}
where $\lambda$ is the spin-orbit coupling constant, $\vec l^{mn}_{i}$ is the matrix element of the orbital angular momentum between
the {\it m}th excited state and the {\it n}th ground state Wannier functions which are centered at {\it i}th ion,
while $\epsilon^{n}_{i}$ represents the energy of the {\it n}th Wannier orbital at the {\it i}th ion.

The second contribution to anisotropic exchange interactions is of Coulomb interaction origin,
\cite{Moriya} and is given by
\begin{eqnarray}
\vec D_{ij}^{Coulomb} = 4i \lambda (\sum_{m} \frac{\vec l^{mn}_{i}}{\epsilon_{i}^{m}-\epsilon_{i}^{n}} 
J_{ij}^{nn'n'm}  \nonumber \\ - \sum_{m'} \frac{\vec l^{m'n'}_{j}}{\epsilon_{j}^{m'}-\epsilon_{j}^{n'}} J_{ij}^{nn'm'n)}), 
\end{eqnarray}
where $J_{ij}^{nn'n'm} = \int \int \frac{W^{n*}_{i}(x) W^{n'}_{j}(x)
W^{n'*}_{j} (x') W^{m}_{i} (x')}{|x-x'|} dx dx'$ is a kind of inter-site Coulomb exchange interaction which is non-diagonal with respect to the orbitals.
Assuming that the Wannier functions of $x^2-y^2$ symmetry are ground state orbitals, these intersite Coulomb exchange interaction integrals can be expressed in the atomic wave function basis in the following form:
\begin{eqnarray}
\int \int \frac{\phi^{*}_{p_y}(x) \phi_{p_x}(x) 
\phi^{*}_{p_x}(x') \phi_{p_z} (x')}{|x-x'|} dx dx' 
\end{eqnarray}
and 
\begin{eqnarray}
\int \int \frac{\phi^{*}_{p_x}(x) \phi_{p_z}(x) 
\phi^{*}_{p_y}(x') \phi_{p_x} (x')}{|x-x'|} dx dx'. 
\end{eqnarray}
One can estimate these integrals through complex spherical harmonics and Slater integrals. \cite{slater}
We found that, due to symmetry, both integrals of Eq.(8) and Eq.(9) are identically zero.
Therefore, we only consider the kinetic Dzyaloshinskii-Moriya interaction in the following.

To perform the microscopic analysis and calculate the kinetic Dzyaloshinskii-Moriya interactions,
we define the hopping parameters of the following general tight-binding 
Hamiltonian that includes five Wannier orbitals centered at Cu sites:
\begin{eqnarray}
H^{5 \, orb}_{TB} = \sum_{\substack {i, j, \sigma \\ k, k'}} t^{kk'}_{ij} a^{+}_{i k \sigma} a_{j k' \sigma},
\end{eqnarray}
where $k,k'=xy, yz, 3z^2-r^2, xz, x^2-y^2$.
Using the projection procedure \cite{proj} we have calculated the hopping integrals between 
the ground state Wannier orbitals of $x^2-y^2$ symmetry
t$_{12}^{x^2-y^2}$= 161 meV, t$_{13}^{x^2-y^2}$ = t$_{14}^{x^2-y^2}$= t$_{15}^{x^2-y^2}$= t$_{16}^{x^2-y^2}$ = 62 meV, t$_{17}^{x^2-y^2}$=t$_{18}^{x^2-y^2}$=16 meV, t$_{19}^{x^2-y^2}$ = t$_{1 \, 10}^{x^2-y^2}$= t$_{1 \, 11}^{x^2-y^2}$= t$_{1 \, 12}^{x^2-y^2}$ = 13 meV, t$_{1A}^{x^2-y^2}$=t$_{1B}^{x^2-y^2}$=12 meV and
t$_{1C}^{x^2-y^2}$ = t$_{1 D}^{x^2-y^2}$= t$_{1F}^{x^2-y^2}$= t$_{1E}^{x^2-y^2}$ = 10 meV.
The hopping integrals between the ground state ($x^2-y^2$) and excited ($xy$, $yz$, $3z^2-r^2$, xz) states Wannier orbitals are given 
\begin{eqnarray}
t_{12}^{yz \, x^2-y^2}=-t_{12}^{x^2-y^2 \, yz}=-t_{12}^{xz \, x^2-y^2} = t_{12}^{x^2-y^2 \, xz} 
\nonumber \\
=-t_{21}^{yz \, x^2-y^2}= t_{21}^{x^2-y^2 \, yz}=t_{21}^{xz \, x^2-y^2} = - t_{21}^{x^2-y^2 \, xz}\nonumber \\  = 10 \, meV \nonumber 
\end{eqnarray}
and
\begin{eqnarray}
t_{13}^{x^2-y^2 \, xy} =  t_{31}^{xy \, x^2-y^2} =  t_{14}^{xy \, x^2-y^2} = t_{41}^{x^2-y^2 \, xy}  \nonumber \\
=- t_{15}^{xy \, x^2-y^2}= -t_{51}^{x^2-y^2 \, xy} = -t_{16}^{x^2-y^2 \, xy} = -t_{61}^{xy \, x^2-y^2} \nonumber \\
= -22 \, meV, \nonumber
\end{eqnarray}
\begin{eqnarray}
t_{13}^{xy \, x^2-y^2} = t_{31}^{x^2-y^2 \, xy} = t_{14}^{x^2-y^2 \, xy} = t_{41}^{xy \, x^2-y^2} \nonumber \\
=- t_{15}^{x^2-y^2 \, xy} = -t_{51}^{xy \, x^2-y^2} = -t_{16}^{xy \, x^2-y^2} = - t_{61}^{x^2-y^2 \, xy} \nonumber \\
= - 12 \, meV, \nonumber
\end{eqnarray}
\begin{eqnarray}
t_{13}^{x^2-y^2 \, xz} = t_{31}^{xz \, x^2-y^2} = t_{14}^{yz \, x^2-y^2} = t_{41}^{x^2-y^2 \, yz} \nonumber \\
=- t_{15}^{xz \, x^2-y^2} = -t_{51}^{x^2-y^2 \, xz} = -t_{16}^{x^2-y^2 \, yz} = -t_{61}^{yz \, x^2-y^2} \nonumber \\
= 15 \, meV. \nonumber
\end{eqnarray}
In contrast to the one-orbital model, the value of the intra-dimer hopping integral between Wannier orbitals of $x^2-y^2$ symmetry becomes smaller. 
This is due to the fact that the hybridization between filled and vacant orbitals is 
explicitly taken into account in the five-orbital model. We can estimate the isotropic 
exchange interactions in the case of the five-orbital model using  Eq.(3),
$J_{12}$ = 7.8 meV, $J_{13}$=3.5 meV, $J_{17}$ = 0.23 meV, $J_{19}$ = 0.15 meV, $J_{1A}$ = 0.13 meV and $J_{1C}$ = 0.09 meV. These values are in better agreement with the experimental estimates.

The energies of the Wannier orbitals obtained by the projection 
procedure \cite{proj} are shown in Table I. 
\begin{table}[!h]
\centering
\caption [Bset]{The energies, $\epsilon_m$ of the Wannier orbitals obtained using the projection procedure (in eV).
The zero energy corresponds to the Wannier orbitals of $3z^2-r^2$ symmetry.}
\label {basisset}
\begin {tabular}{ccccc}
  \hline
  \hline
          $xy$   & $yz$  & $3z^2-r^2$  & $zx$  & $x^2-y^2$   \\
  \hline
          0.26   & 0.20  &  0          & 0.20 &  2.54          \\
  \hline
  \hline
\end {tabular}
\end {table}
Using these hopping integrals, we find that Eq.(5) can be reduced to the following expression 
for the $x$ and $y$ components of the intradimer Dzyaloshinskii-Moriya interaction (i=1 and j=2):
\begin{eqnarray}
D^{x}_{12} = \frac{16 \lambda}{\alpha^4 U_{d} (\epsilon_{yz} - \epsilon_{x^2-y^2})} t^{x^2-y^2}_{12} t^{yz \, x^2-y^2}_{21}
\end{eqnarray} 
and 
\begin{eqnarray}
D^{y}_{12} = \frac{16 \lambda}{\alpha^4 U_{d} (\epsilon_{xz} - \epsilon_{x^2-y^2})} t^{x^2-y^2}_{12} t^{xz \, x^2-y^2}_{21}.
\end{eqnarray}
The z component of the inter-dimer (i=1 and j=3,4,5,6) coupling is given by
\begin{eqnarray}
D^z_{ij} = \frac{8 \lambda t^{x^2-y^2}_{ij}}{\alpha^4 U_{d} (\epsilon_{xy} - \epsilon_{x^2-y^2})} \nonumber \\ 
\times [t_{ji}^{x^2-y^2 \, xy} - t_{ji}^{xy \, x^2-y^2} + t_{ij}^{xy \, x^2-y^2} - t_{ij}^{x^2-y^2 \, xy}].  
\end{eqnarray}
and the x and y components of the inter-dimer interactions are given by
\begin{eqnarray}
D^y_{13} = \frac{4 \lambda t^{x^2-y^2}_{13}}{\alpha^4 U_{d} (\epsilon_{xz} - \epsilon_{x^2-y^2})} [t_{31}^{xz \, x^2-y^2} + t_{13}^{x^2-y^2 \, xz}],
\end{eqnarray}
\begin{eqnarray}
D^x_{14} = \frac{4 \lambda t^{x^2-y^2}_{14}}{\alpha^4 U_{d} (\epsilon_{yz} - \epsilon_{x^2-y^2})} [t_{41}^{x^2-y^2 \, yz} + t_{14}^{yz \, x^2-y^2}],
\end{eqnarray}
\begin{eqnarray}
D^y_{15} = \frac{4 \lambda t^{x^2-y^2}_{15}}{\alpha^4 U_{d} (\epsilon_{xz} - \epsilon_{x^2-y^2})} [t_{51}^{x^2-y^2 \, xz} + t_{15}^{xz \, x^2-y^2}],
\end{eqnarray}
and 
\begin{eqnarray}
D^x_{16} = \frac{4 \lambda t^{x^2-y^2}_{16}}{\alpha^4 U_{d} (\epsilon_{yz} - \epsilon_{x^2-y^2})} [t_{61}^{yz \, x^2-y^2} + t_{16}^{x^2-y^2 \, yz}].
\end{eqnarray}

We are now in a position to calculate Dzyaloshinskii-Moriya interactions using Eq.(11-17) and the value of spin-orbit coupling $\lambda$ = 0.1 eV. These results are presented in Table II and Fig.8. In the notation of Eq.(1) $\vec D$ and $\vec D'$ correspond to $\vec D_{12}$ and $\vec D_{1j}$ (where j=3,4,5,6), respectively.
\begin{table}[!h]
\centering
\caption [Bset]{Calculated anisotropic exchange interaction vectors $\vec D_{ij}$ (in meV).}
\label {basisset}
\begin {tabular}{cc}
  \hline
  \hline
  (i,j)              & $\vec D_{ij} $   \\
  \hline
  (1,2)           &  (0.25;-0.25;0.0)    \\
  (1,3)           &  (0.0; -0.07 ; 0.1)      \\
  (1,4)           &  (-0.07; 0.0; -0.1)      \\
  (1,5)           &  (0.0; 0.07; 0.1)       \\
  (1,6)           &  (0.07;0.0;-0.1)       \\
  \hline
  \hline
\end {tabular}
\end {table}
Based on Eq.(11-12), we can conclude that the source of intra-dimer anisotropic exchange 
interaction is the hopping process between Wannier orbitals of $x^2-y^2$ and $yz (xz)$ symmetry.
The microscopic origin of the inter-dimer Dzyaloshinskii-Moriya interaction is the transfer process 
between $x^2-y^2$ and $xy$ Wannier orbitals.
\begin{figure}[!b]
\centering
\includegraphics[angle=0,width=0.3\textwidth]{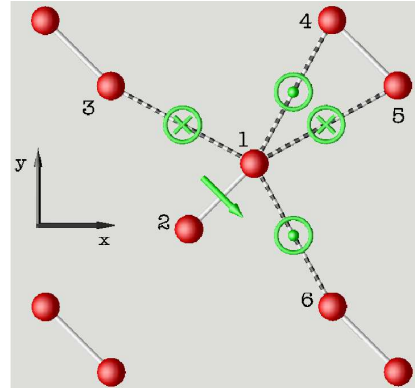}
\caption {Schematic representation of symmetry of Dzyaloshinskii-Moriya interactions.}
\end{figure}
It is interesting to note that there are $x$ and $y$ components of inter-dimer anisotropic 
couplings. 
The comparison with experimental data
is presented in Sec.VI.
In the next section we present the results of LSDA+U calculations, which give additional confirmation of our model consideration for the isotropic interactions.

\section{Results of LSDA+U calculations}
The results of LDA calculations have demonstrated that the magnetic properties of SrCu$_{2}$(BO$_{3}$)$_{2}$ can be 
reproduced correctly within the microscopic model approach based on the LDA calculations. 
However, LDA cannot correctly describe the insulating ground state of SrCu$_2$(BO$_3$)$_2$. 
In order to overcome this problem, we have used the LSDA+U approximation \cite{Anisimov} which 
takes into account the Coulomb correlations of localized states neglected in LDA. 
The effective Coulomb interaction U$_{d}$ and the effective intra-atomic exchange J$^{H}_{d}$ of the copper atoms, which represent external parameters in a 
self-consistent cycle of the LSDA+U scheme, are determined from the first-principle calculation by constrained LDA. 
The calculation scheme has been described elsewhere.\cite{gunn}
The Coulomb interaction parameter $\tilde U_{d}$ and the intra-atomic exchange $\tilde J^{H}_{d}$ have been estimated to be $\tilde U_{d}$=9.4 eV and $\tilde J^{H}_{d}$=1 eV.

We have performed LSDA+U calculations for two magnetic configurations which are presented in Fig.9. 
The results of these calculations are presented in Table III. 
In both cases, SrCu$_{2}$(BO$_{3}$)$_{2}$ is an insulator with an energy gap of 2.2 eV.
The antiferromagnetic configuration has a lower energy.

Let us describe the inter-dimer interactions. For that we use the Green's function method.
Following Lichtenstein {\it et al.} \cite{Liechtenstein}, we determine the exchange interaction
parameter between copper atoms
via the second variation of the total energy with respect to small deviations of the magnetic moments
from the collinear magnetic configuration.
The exchange interaction parameters J$_{ij}$
can be written in the following form: \cite{Liechtenstein,Mazurenko}
\begin{eqnarray}
J_{ij} = \frac{2}{\pi} \int_{-\infty}^{E_{F}} d\epsilon \, {\rm Im} \sum_{\substack {m, m' \\ m'', m'''}}
(\Delta^{mm'}_{i} \,
G_{ij \, \downarrow}^{m'm''} \, \Delta^{m'' m'''}_{j} \, G_{ji \, \uparrow}^{m''' m}), \nonumber
\end{eqnarray}
where $m$ is the magnetic quantum number, the on-site potential $\Delta^{mm'}_{i}=H^{m m'}_{ii \, \uparrow} - H^{m m'}_{ii \, \downarrow}$
and the Green's function is calculated in the following way
\begin{eqnarray}
G^{mm'}_{ij \sigma}(\epsilon) \, = \, \sum_{k,\, n} \frac{c^{mn}_{i \sigma} \, (k) \, c^{m'n *}_{j \sigma} \,
(k)}{\epsilon-E^{n}_{\sigma}}.
\end{eqnarray}
Here $c^{mn}_{i\sigma}$ is a component of the {\it n}th eigenstate, and
E$_{\sigma}^{n}$ is the corresponding eigenvalue.
This leads to the following inter-dimer exchange interactions: $J_{13}$=$J_{14}$=$J_{15}$=$J_{16}$=4.1 meV, $J_{17}$=$J_{18}$=0.17 meV, $J_{1 \, 9}$=$J_{1 \, 10}$ = $J_{1 \, 11}$ = $J_{1 \, 12}$ = 0.06 meV, $J_{1A}$=$J_{1B}$ = 0.06 meV, $J_{1C}$ = $J_{1D}$ = $J_{1E}$=$J_{1F}$= 0.02 meV.

One can see that in the case of the antiferromagnetic configuration there is no magnetic moment at the oxygen atom (Table III). But in the ferromagnetic configuration, the oxygen has a small moment. This fact supports the scenario proposed in Ref. [\onlinecite{mazurenkoLiCu2O2}]. Therefore, to calculate the intra-dimer exchange interaction, it
is necessary to take into account the change of oxygen magnetization. One can do this using the method of total energies difference. 
\begin{figure}[h]
\centering
\includegraphics[angle=0,width=0.4\textwidth]{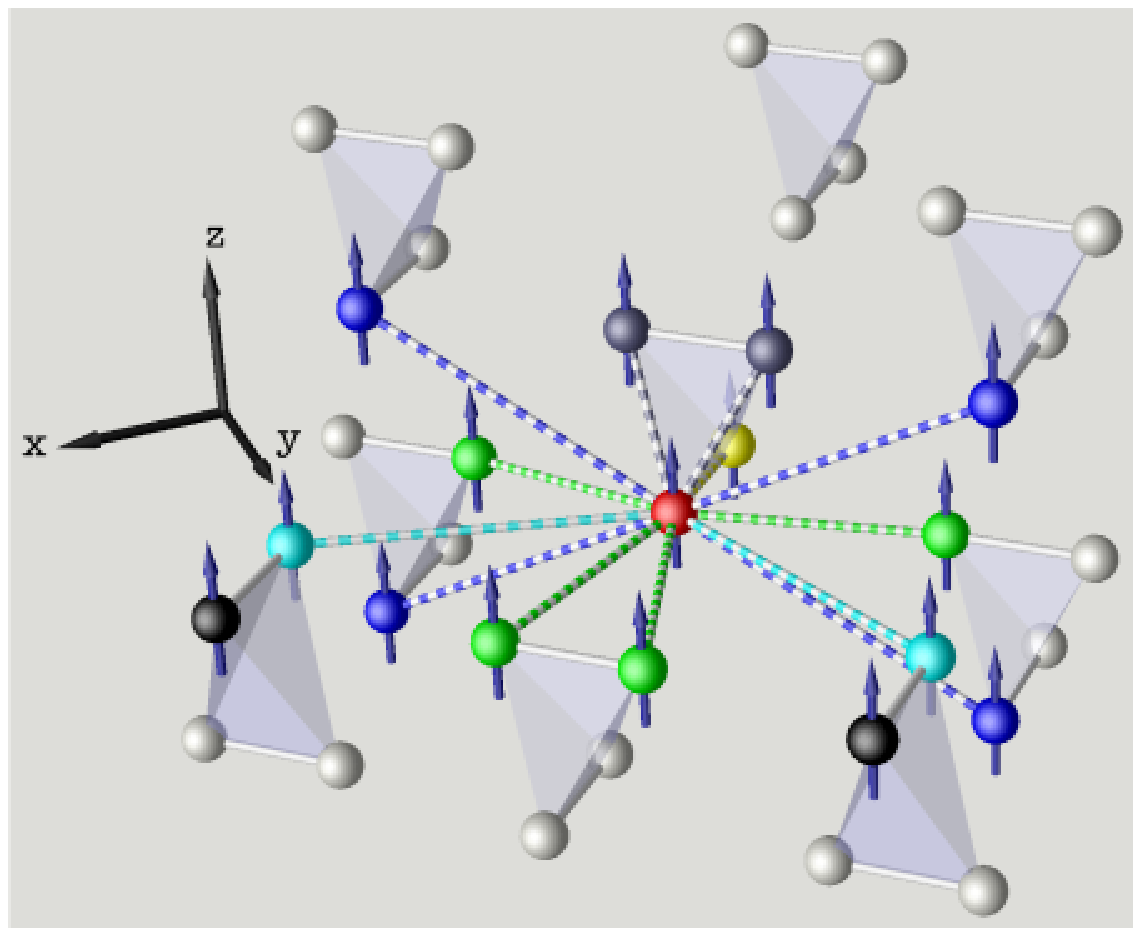}
\includegraphics[angle=0,width=0.4\textwidth]{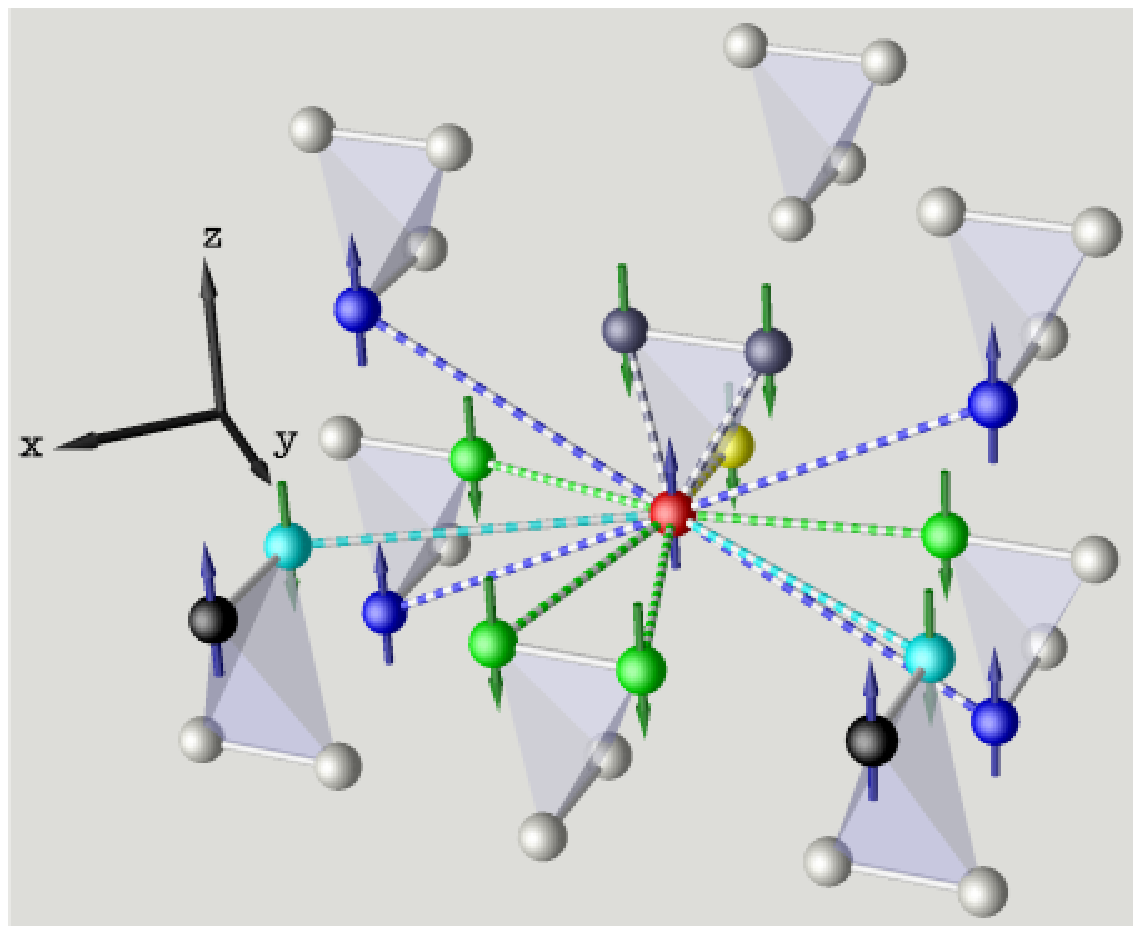}
\caption {Ferromagnetic and antiferromagnetic configurations simulated in LSDA+U calculations.}
\end{figure}
\begin{table}
\centering
\caption [Bset]{Results of LSDA+U calculations for ferromagnetic and antiferromagnetic configurations. 
E$_{gap}$ is the value of the energy gap (in eV). M(Cu) and M(O) are the values of the magnetic moments of copper and oxygen atoms which are located 
between the copper atoms in dimer (in $\mu_{B}$).
E$_{tot}$ is the relative total energy of the system in different magnetic configurations (in meV).}
\label {basisset}
\begin {tabular}{ccccc}
  \hline
  \hline
                & E$_{gap}$   & M(Cu)   &  M(O)       & E$_{total}$  \\
  \hline
  FM            &   2.2       &   0.72          &  0.1      &  12  \\
  AFM           &   2.2       &   0.72          &  0         &  0      \\
  \hline
  \hline
\end {tabular}
\end {table}
For that purpose, we write the Heisenberg Hamiltonian in the following form:
\begin{equation}
H=\sum_{i>j} J_{ij} \vec {S_{i}} \vec {S_{j}}.
\end{equation}
The total energies of the ferromagnetic and antiferromagnetic configurations 
presented in Fig.9 are
\begin{equation}
E_{FM}=   J_{12} S^2 + 4 J_{13} S^2 + 2 J_{17} S^2
\end{equation}
and 
\begin{equation}
E_{AFM}= -  J_{12} S^2 - 4 J_{13} S^2 - 2 J_{17} S^2.
\end{equation}
Therefore, the dimer exchange interaction $J_{12}$ is given by
\begin{equation}
J_{12} = \frac{E_{FM}-E_{AFM}-8 J_{13} S^2 - 4 J_{17} S^2}{2 S^2}.
\end{equation}
Using calculated values of the inter-dimer couplings ($J_{13}$ = 4.1 meV and $J_{17}$ = 0.17 meV) and the values of the total energies from Table III, one can calculate the dimer exchange interaction for S=1/2, which leads to $J_{12}$ = 7.3 meV.

\section{Comparison with experiment}

The exchange interactions obtained with different methods are summarized in Table IV, together
with the values most often used to explain experimental data. The Green's function method
is in principle the most accurate, and if its results are in good agreement with those obtained
by LDA using microscopic exchange formulae, which is the case here, the results are expected
to be quite accurate. The agreement with experimental estimates is quite good for $J_{12}$ and $J_{13}$.
The inter-plane coupling $J_{1A}$ is however predicted to be about 10 times smaller than what
has been suggested from fitting the susceptibility, a conclusion to be taken seriously in view
of the overall good agreement regarding the other parameters. Besides, the dominant couplings 
beyond $J_{12}$ and $J_{13}$ are predicted to be the in-plane exchange couplings $J_{17}$ and $J_{19}$,
parameters which have not been considered so far in theoretical models of SrCu$_2$(BO$_3$)$_2$.

Let us go beyond these general statements and look in more details at how well the 
calculated exchange parameters can reproduce using  basic experimental facts such as 
the spin gap and the temperature dependences of the magnetic susceptibility and specific heat.
Let us start with the spin gap analysis. According to different experiments \cite{Miyahara}, the value of the spin gap varies from 2.6 meV (nuclear quadrupole resonance) to 3.1 meV (nuclear magnetic resonance).
Based on the calculated exchange interactions (Table IV), if we estimate the spin gap using the following expression \cite{Miyahara}
\begin{eqnarray}
\Delta= J_{12}(1-(\frac{J_{13}}{J_{12}})^2 - \frac{1}{2} (\frac{J_{13}}{J_{12}})^3 - \frac{1}{8}(\frac{J_{13}}{J_{12}})^4 ). 
\label{gap}
\end{eqnarray}
the spin gap value is 9.1 meV in the case of the one-orbital model.
This is about three times larger than experimental value. A better agreement is achieved in the five-orbital model and LSDA+U calculations, $\Delta^{5orb}$=5.8 meV and $\Delta^{LSDA+U}$=4.2 meV.
Let us note however that $J_{13}/J_{12}\simeq0.6$ is not that small, and higher order corrections 
to the estimate of Eq.(\ref{gap}) are expected to further reduce the gap, hence to improve the
agreement with the experimental determination of the gap.

The results of measurements of magnetic susceptibility and specific heat can be reproduced
using the Heisenberg model with different sets of isotropic parameters. For instance, in the review of Miyahara and Ueda \cite{Miyahara}, there are five different sets of exchange couplings which vary from $J$ = 6.6 meV and $J'$ = 4.1 meV
to $J$ = 7.3 meV and $J'$ = 4.6 meV.
The latter $J$ and $J'$ correspond to the best fitting to the experimental data. These values are in reasonable agreement with our exchange
interactions for the five-orbital model ($J_{12}$ = 7.8 meV and $J_{13}$= 3.5 meV) and in good agreement with those obtained by LSDA+U calculations ($J_{12}$ = 7.3 meV and $J_{13}$= 4.1 meV)

\begin{table}
\centering
\caption [Bset]{Values of exchange interactions J$_{ij}$ between magnetic moments of 
SrCu$_{2}$(BO$_{3}$)$_{2}$ system (in meV).}
\label {basisset}
\begin {tabular}{ccccc}
  \hline
  \hline
           & LDA$^a$   & LDA$^b$   & LDA+U$^c$    &  Exp. (after Ref. \onlinecite{Miyahara} )        \\
  \hline
  J$_{12}$        & 10.6    & 7.8      &  7.3      &  7.3            \\
  J$_{13}$        & 3.6     & 3.5      &  4.1      &  4.6          \\
  J$_{17}$        & 0.2     & 0.23     &  0.17     &   -         \\
  J$_{19}$        & 0.13    & 0.15     &  0.06     &   -         \\
  J$_{1A}$        & 0.06    & 0.13     &  0.06     &  0.66         \\
  J$_{1C}$        & 0.13    & 0.09     &  0.02     &   -         \\
  \hline
  \hline
\end {tabular}

$^a$ One-orbital model. \\
$^b$ Five-orbital model. \\
$^c$ Green's function approach.
\end {table}

The presence of a finite magnetization well below the expected critical field for the gap closing is a manifestation of the fact that triplet states are mixed into the ground state. This is an effect of the Dzyaloshinskii-Moriya interaction.
The directions (Fig.8) and values (Table II) of calculated Dzyaloshinskii-Moriya vectors agree well with those obtained from the interpretation of neutron, ESR and NMR experiments.\cite{Miyahara,Frederic}
For instance, the intra-dimer anisotropic vector is perpendicular to the bond direction.
The inter-dimer Dzyloshinskii-Moriya interactions
lie mainly along the z axis. The calculated ratios $|\vec D| / J$ = 0.032 and $|\vec D'| / J$ = 0.016 (with J = 7.8 meV) are in good agreement with those deduced from experiments ($|\vec D| / J$ = 0.034 and $|\vec D'| / J$ = 0.02).

\section{CONCLUSION}
In this paper we have presented the results of an ab-initio investigation of the magnetic properties of SrCu$_{2}$(BO$_{3}$)$_{2}$.
It is found that the ferromagnetic contribution of the intra-atomic exchange interactions of oxygen plays a
crucial role to account for the
intra-dimer isotropic exchange interaction.
The microscopic origin of Dzyaloshinskii-Moriya interactions has been analyzed using calculated 
hopping integrals between different Wannier orbitals of copper atoms. 
In general, there are two contributions to anisotropic exchange interaction which are kinetic and Coulomb Dzyaloshinskii-Moriya interactions. In the case of SrCu$_{2}$(BO$_{3}$)$_{2}$ the Coulomb contribution is zero by symmetry. However,
we expect that the latter plays important role in the case of more distorted frustrated cuprates such as
Cu$_2$Te$_2$O$_5$(Br,Cl)$_2$ and (Tl,K)CuCl$_3$.
  
\section{ACKNOWLEDGMENTS}
We would like to thank S. Miyahara, A. Gell\'e, A.I. Lichtenstein, M. Sigrist, M. Troyer, A.O. Shorikov, F. Lechermann for helpful discussions and S.V. Streltsov for his assistance with LDA calculations. 
The hospitality of the Institute of Theoretical Physics of EPFL is gratefully acknowledged.
This work is supported by INTAS Young Scientist Fellowship Program Ref. Nr. 04-83-3230, 
Netherlands Organization for Scientific Research 
through NWO 047.016.005, 
Russian Foundation for Basic Research grant RFFI 07-02-00041, RFFI 06-02-81017 and the grant program of President of Russian Federation
Nr. MK-1041.2007.2.
The calculations have been performed by the computer cluster of ``University Center of Parallel Computing'' of USTU-UPI. 
We also acknowledge the financial support of the Swiss National Fund and of MaNEP.

\end{document}